\documentclass[11pt]{article}

\usepackage[margin=1in]{geometry}
\usepackage[T1]{fontenc}
\usepackage[utf8]{inputenc}
\usepackage{lmodern}
\usepackage{microtype}
\usepackage{amsmath,amssymb,mathtools}
\usepackage{graphicx}
\usepackage{booktabs,array,tabularx}
\usepackage{siunitx}
\usepackage{algorithm}
\usepackage{algpseudocode}
\usepackage[numbers,sort&compress]{natbib}
\usepackage{tikz}
\usetikzlibrary{arrows.meta,positioning,fit,calc,backgrounds}
\usepackage{hyperref}
\usepackage[nameinlink,noabbrev]{cleveref}
\usepackage{enumitem}

\newcommand{\method}{\textsc{Sec2Drum-DAC}}
\newcommand{\R}{\mathbb{R}}
\newcommand{\E}{\mathbb{E}}
\newcommand{\given}{\,|\,}
\newcommand{\fadinf}{\ensuremath{\mathrm{FAD}_{\infty}}}
\newcommand{\fadrtwo}{\ensuremath{\mathrm{FAD}\text{-}R^2}}

\hypersetup{
  colorlinks=true,
  linkcolor=blue,
  citecolor=blue,
  urlcolor=blue
}

\setlist{nosep,leftmargin=1.15em}

\title{Seconds-Aligned PCA-DAC Latent Diffusion for Symbolic-to-Audio Drum Rendering}
\author{Anonymous Author(s)}

\author{
Konstantinos Soiledis\textsuperscript{1,2}
\and
Maximos Kaliakatsos-Papakostas\textsuperscript{1}
\and
Dimos Makris\textsuperscript{1}
\and
Konstantinos Tsamis\textsuperscript{1,2}
\\[0.75em]
\small \textsuperscript{1}Dept. of Music Technology and Acoustics, Hellenic Mediterranean University, Rethymno \& Athens, Greece
\\
\small \textsuperscript{2}Athena RC, Athens, Greece
\\
\small \texttt{k.t.soiledis@gmail.com}
}

\date{}

\begin{document}
\maketitle

\begin{abstract}
Symbolic-control drum generation requires preserving explicit event timing and dynamics while synthesizing acoustically plausible waveforms. We present \method{}, a conditional latent-diffusion model for symbolic-to-audio drum rendering. The model conditions on event features sampled in physical time at codec-frame locations and predicts standardized principal-component coordinates of frozen DAC summed-codebook embeddings rather than waveform samples. In the evaluated DAC configuration, 72 principal components capture the observed training-frame summed-latent subspace under the stated SVD threshold, yielding a compact continuous denoising target with a deterministic reconstruction path to the 1024-dimensional DAC latent space before waveform decoding.

Across 1{,}733 held-out four-beat windows, PCA diffusion improves paired spectral and transient metrics over deterministic PCA regression and a symbolic rendering baseline, while direct regression remains stronger on phase-sensitive waveform $L_1$. Auxiliary RVQ cross-entropy improves short-step diffusion on mel error, onset-flux cosine, and waveform $L_1$, with the most favorable trade-offs occurring at 6--25 denoising steps depending on the metric.
\end{abstract}

\section{Introduction}

Controllable drum synthesis requires more than plausible percussion generation: it must realize a prescribed event pattern with high temporal and timbral fidelity. Symbolic representations expose musically meaningful variables such as onset timing, velocity, and instrument family, and therefore support precise editing and structural control. However, symbolic systems often culminate in MIDI playback or sample triggering, which constrains acoustic diversity. Neural audio generators can operate in a richer acoustic space, but conditioning may be underspecified: a control sequence can indicate intended events without ensuring that the resulting waveform faithfully realizes them. This mismatch is particularly important when the objective is scientific evaluation of controllability rather than unconstrained generative quality.

We therefore study symbolic-to-audio drum rendering. Let $c$ denote a symbolic drum-control sequence and let $a \in \R^L$ denote the corresponding waveform segment. The task is to learn a conditional generator $p_\theta(a \given c)$ that preserves the rhythmic, dynamic, and family-level structure encoded in $c$ while producing realistic drum audio. The four-beat window setting permits analysis of representation choice and control fidelity against paired audio targets.

Within this problem, the central modeling question is how to choose a target representation that is both decodable and tightly coupled to symbolic control. \method{} addresses this question through three design choices. First, symbolic conditioning is aligned to codec-frame times in physical seconds rather than to a shared symbolic index. Second, the denoising target is a continuous PCA coordinate representation of DAC summed-codebook embeddings rather than waveform samples. Third, the model optionally incorporates an auxiliary RVQ cross-entropy objective that regularizes continuous predictions with respect to the codec's residual-quantization structure. These choices define the following generation pipeline:
\[
\text{drum grid} \rightarrow \text{seconds-aligned conditioning}
\rightarrow \text{PCA-DAC latent diffusion}
\rightarrow \text{DAC decoding}.
\]

Alongside the artifact-backed evaluation, we maintain an interactive Gradio listener interface for qualitative inspection \citep{abid2019gradio}; the local environment uses Gradio 5.35.0. The interface edits a 16-step drum sketch and renders conditioning grids and audio previews from the same checkpoint family, but it is treated as demo infrastructure rather than as user-study evidence.

The contributions are fourfold:
\begin{enumerate}
  \item We formulate symbolic-to-audio drum rendering at codec-frame resolution with conditioning features computed in physical time.
  \item We introduce a continuous PCA-DAC latent target that reduces denoising dimensionality while retaining a deterministic reconstruction path to full DAC latent space.
  \item We study auxiliary RVQ cross-entropy as codec-aware regularization for continuous latent denoising.
  \item We report a standardized artifact-backed evaluation comparing symbolic rendering, deterministic regression, and latent-diffusion variants.
\end{enumerate}

\section{Related Work}

\paragraph{Diffusion and latent diffusion.}
Denoising diffusion probabilistic models established diffusion as a practical framework for high-fidelity likelihood-based generation \citep{ho2020ddpm}, and subsequent work improved sample quality and scalability in large-scale image synthesis \citep{dhariwal2021diffusion}. Latent diffusion showed that denoising can be performed in compressed autoencoder spaces, substantially reducing computational cost while retaining perceptual quality \citep{rombach2022ldm}. Diffusion autoencoders further demonstrated that diffusion models can be paired with structured and decodable latent representations rather than only pixel-space sampling \citep{preechakul2022diffae}. Transformer denoisers such as DiT and PixArt-$\alpha$ have established a block-level convention in which latent tokens are denoised by repeated attention/feed-forward blocks with timestep modulation and, for conditional generation, cross-attention into the conditioning sequence \citep{peebles2023dit,chen2024pixartalpha}. The present work adopts this latent-denoising perspective in the setting of explicitly conditioned drum rendering rather than unconstrained image synthesis.

\paragraph{Codec-based audio generation.}
Recent audio generation systems increasingly rely on learned low-rate intermediate representations. SoundStream, EnCodec, and DAC establish neural-codec representations with residual vector quantization and high-fidelity decoding \citep{zeghidour2022soundstream,defossez2022encodec,kumar2023dac}. AudioLM and MusicLM showed that neural-codec representations can support broad audio and music generation \citep{borsos2023audiolm,agostinelli2023musiclm}. AudioLDM~2 demonstrates diffusion over learned audio representations for broad audio-generation tasks \citep{liu2024audioldm2}, while Stable Audio and long-form music diffusion show that timing-aware latent diffusion can scale to long-duration, high-resolution synthesis \citep{evans2024stableaudio,evans2024longform}. These systems establish the viability of learned audio representations, but they are not designed specifically for event-faithful rendering from an explicit drum-control grid.

\paragraph{Symbolic drums and drum audio control.}
Prior work on symbolic drums emphasizes controllability, groove, and expressive timing, including sequence and performance modeling on the Groove MIDI Dataset \citep{gillick2019learning,magenta2019groove}. DrumGAN targets isolated drum-sound synthesis with timbral conditioning rather than full groove rendering \citep{nistal2020drumgan}, and CRASH uses score-based raw-audio generative modeling for controllable high-resolution drum sound synthesis \citep{rouard2021crash}. More recent diffusion-based approaches generate symbolic drum patterns from text and therefore treat the drum sequence itself as the output object \citep{jajoria2024text}. The present setting reverses this direction: the symbolic drum grid is observed, and the modeling objective is to render waveform audio that preserves rhythmic, dynamic, and family-level structure.

\paragraph{Symbolic-to-audio rendering.}
Symbolic-to-audio synthesis has been explored for general music and tonal instruments, including spectrogram-based multi-instrument synthesis \citep{hawthorne2022multiinstrument} and MIDI-to-piano synthesis with neural codec language modeling \citep{tang2025midivalle}. A closely related recent preprint is the expressive drum-grid codec-token study of \citet{soiledis2026drumsynthesis}: it maps time-aligned drum grids with microtiming and velocity information to discrete codec IDs and compares EnCodec, DAC, and X-Codec on E-GMD. The present work retains the explicit drum-grid rendering task and paired objective-metric evaluation, but changes the modeling target to continuous PCA-DAC summed-latent diffusion. Discrete RVQ supervision appears only as an auxiliary training regularizer, not as the primary output representation.

\paragraph{MIDI-to-drum and prompt-controlled drum rendering.}
The most directly comparable published MIDI-to-drum audio system we found is Break-the-Beat!: it fine-tunes Stable Audio Open for drum MIDI plus reference-audio timbre control, represents MIDI as tempo-relative arrangement and tap grids, and evaluates audio quality, rhythmic alignment, and beat continuity \citep{cui2026breakthebeat}. Its control problem is therefore timbre-conditioned drum rendering from MIDI plus a reference audio example. Audio-prompted drum systems such as TRIA and DARC also target controllable drum generation, but their control interface is a rhythmic audio prompt such as tapping or beatboxing, often combined with timbre or musical-context prompts \citep{oreilly2025tria,brosnan2026darc}. In contrast, \method{} does not perform reference-timbre transfer or adapt a text-to-audio backbone; it evaluates paired rendering from an explicit velocity/articulation drum grid into a continuous DAC-latent trajectory aligned in physical seconds.

\section{Method}

\subsection{Problem Formulation}

Each training example consists of a symbolic control sequence $c=\{c_t\}_{t=1}^{T_c}$ and a paired waveform segment $a \in \R^L$. Let $E_{\mathrm{enc}}$, $Q_{\mathrm{RVQ}}$, and $D_{\mathrm{DAC}}$ denote the frozen DAC encoder, residual-vector quantizer, and decoder. The encoder first maps the waveform to a pre-quantizer representation $u=E_{\mathrm{enc}}(a)$. The RVQ stack then selects code indices and their projected codebook embeddings,
\[
(q_{1:T},y_{1:T})=Q_{\mathrm{RVQ}}(u), \qquad
y_j=\sum_{k=1}^{K_{\mathrm{rvq}}} e_{k,q_{j,k}}, \qquad y_j \in \R^D,
\]
with $D=1024$ and $K_{\mathrm{rvq}}=9$ in the present configuration. The target is therefore the post-quantizer summed-codebook embedding, not the raw DAC encoder state. Rather than modeling $p_\theta(a \given c)$ directly in waveform space, we learn a conditional model over normalized PCA coordinates of this decodable latent representation and recover waveform audio through the frozen decoder. The induced waveform distribution is therefore mediated by a model over latent trajectories:
\[
p_\theta(\tilde{z}_{1:T} \given c),
\]
which induces waveform samples through PCA reconstruction and the deterministic map $\hat{a}=D_{\mathrm{DAC}}(\hat{y}_{1:T})$. Operationally, the learning problem is reduced to conditional generation of a codec-latent coordinate trajectory whose acoustic realization is fixed once the reconstructed latent is specified.

\subsection{Seconds-Aligned Symbolic Conditioning}

The conditioning representation is an eight-family drum grid with per-family state velocity, onset velocity, and onset count, yielding 24 numeric channels over symbolic time. A key modeling choice is to avoid assuming that symbolic grid indices are commensurate with codec-frame indices. Such an assumption becomes brittle when tempo, symbolic quantization, and codec frame rate interact. Instead, for codec frame $j$ at physical time $\tau_j$, the model computes
\[
h_j = F(c,\tau_j),
\]
where $F$ extracts local symbolic context in seconds. The implemented conditioner uses a hybrid multiscale frontend with radii $0,22,41,55$ grid steps, 64-dimensional features per scale, and a positional encoding defined in physical time. In the reported model, the four scale outputs are concatenated into a 256-channel conditioning vector per codec frame before the denoiser's conditioning projection. Consequently, the denoiser receives event information aligned to codec-frame times rather than to an arbitrary shared index. This seconds-aligned mapping is the main mechanism by which symbolic timing information is made available throughout denoising.

The frontend is a trainable local sequence encoder rather than a fixed resampler. For each codec-frame time, it extracts a token-centered window from the 250~Hz symbolic grid in seconds; numeric lanes are interpolated or sampled according to their feature type, while articulation IDs are expanded into one-hot channels and appended to the local window. Each radius branch uses the same hybrid encoder: a temporal convolutional stem, residual dilated TCN blocks, and a bidirectional LSTM whose center state is layer-normalized and $L_2$-normalized to a 64-dimensional scale feature. The denoiser concatenates the four scale features, projects the resulting 256-dimensional conditioning vector to the Transformer width, and adds the same seconds-based positional encoding used for the noised PCA trajectory.

The frontend configuration used in the reported runs is summarized in \Cref{tab:frontend_config}. Radius 0 denotes the source feature sample at the codec-frame time itself; nonzero radii add symmetric seconds-grid context around that same time.

\begin{table}[t]
\centering
\small
\begin{tabular}{ll}
\toprule
Component & Value \\
\midrule
Symbolic grid rate & 250 Hz \\
Drum families & 8 \\
Numeric channels & 24 \\
Articulation encoding & Per-family IDs expanded to one-hot lanes \\
Radii & 0, 22, 41, 55 grid steps \\
Radii in milliseconds & 0, 88, 164, 220 ms \\
Branch output & 64 \\
Concatenated conditioning dimension & 256 \\
Denoiser width & 768 \\
\bottomrule
\end{tabular}
\caption{Seconds-aligned symbolic frontend configuration for the reported diffusion and direct-regression models.}
\label{tab:frontend_config}
\end{table}

\begin{figure}[t]
\centering
\includegraphics[width=\linewidth]{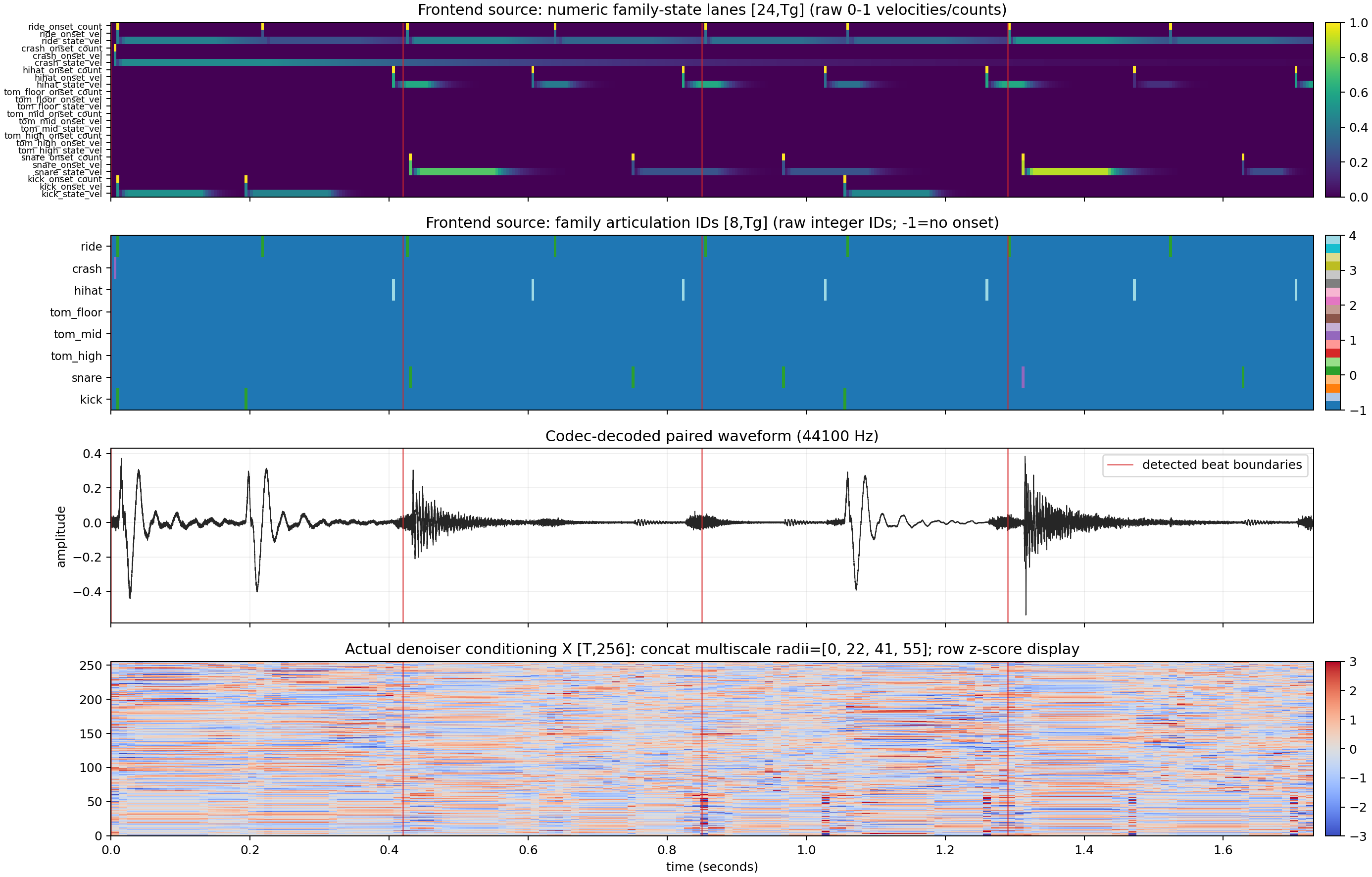}
\caption{Conditioning-side representation inspection for a held-out four-beat window. The raw 24-lane family-state grid and family articulation-ID grid are frontend source fields; they are merged by the seconds-aware frontend into the denoiser conditioning representation $X$ evaluated at codec-frame times. In this configuration, $X$ is the concatenation of four 64-dimensional scale outputs. The plotted $X$ values are row z-scores over time for display only; the model receives the frontend activations directly. No 16th-step aggregate is plotted because it is not a model input. Red lines mark detected beat boundaries.}
\label{fig:conditioning_representations}
\end{figure}

The numeric lanes and per-family articulation IDs are not treated as separate downstream conditions. They are fused by the seconds frontend into a single merged conditioning sequence $X=\{h_j\}_{j=1}^{T}$ evaluated at codec-frame times. \Cref{fig:conditioning_representations} visualizes this process for one held-out four-beat window. The first two panels show the raw source fields consumed by the frontend, not separate denoiser inputs; the actual denoiser-side conditioning is the merged representation $X$ shown in the final panel. The continuous $X$ panel is row-standardized over time only for visualization. Red vertical lines mark the detected beat boundaries used to construct the segment.

\subsection{PCA-Compressed Codec-Latent Target}

The prediction target is the continuous summed DAC codebook-embedding trajectory. We represent this trajectory with PCA fitted on training frames only, thereby avoiding leakage from validation or test audio. Let $\mu \in \R^D$ denote the training-set mean and let $U_K \in \R^{D \times K}$ contain the top $K$ principal directions. For each codec frame,
\begin{equation}
z_j = U_K^\top (y_j-\mu), \qquad z_j \in \R^K .
\end{equation}
The present configuration uses $K=72$. This value is not an unconstrained hyperparameter sweep in the reported cache: the DAC summed-codebook embeddings have recorded subspace rank 72, and the fitted basis accounts for essentially all training-frame variance. The resulting coefficients are standardized using training-set coefficient statistics:
\begin{equation}
\tilde{z}_j = \frac{z_j-m}{s},
\end{equation}
where $m,s \in \R^K$ are the per-dimension mean and standard deviation used by the diffusion model.

This representation is motivated by three considerations: reduced denoising dimensionality relative to the full codec embedding, deterministic reconstruction into full DAC latent dimensionality, and an ordered linear structure that avoids introducing an additional learned bottleneck. PCA-guided diffusion variants have been explored in image generation \citep{song2024pcaddpm}; here, PCA is instead a fixed cache-construction step. The basis is fit once on training-set DAC summed-codebook embeddings, applied unchanged to validation and test frames, and inverted deterministically before DAC decoding.

\Cref{tab:pca_diagnostics} reports diagnostics for the stored PCA basis. The rank claim is an effective rank of the DAC quantizer output-projection subspace, not an unconstrained empirical rank of arbitrary 1024-dimensional vectors. Specifically, the nine DAC quantizer output-projection matrices form a $1024 \times 72$ stack with numerical rank 72 under the tolerance shown below, and the summed codebook embeddings lie in that projected subspace.

\begin{table}[t]
\centering
\small
\begin{tabular}{ll}
\toprule
Diagnostic & Value \\
\midrule
DAC output-projection matrix & $1024 \times 72$ \\
SVD rank tolerance & $7.47\times10^{-4}$ \\
Effective subspace rank & 72 \\
Training frames used for PCA & 1{,}835{,}541 \\
Explained variance, 72 PCs & 100.000000\% \\
Validation latent MSE after PCA inverse & $1.60\times10^{-8}$ \\
Test latent MSE after PCA inverse & $1.60\times10^{-8}$ \\
Test latent RMSE after PCA inverse & $1.27\times10^{-4}$ \\
Test maximum absolute latent error & $6.83\times10^{-4}$ \\
Target PCA reconstruction mel / waveform $L_1$ & 0.10 dB / 0.0002 \\
\bottomrule
\end{tabular}
\caption{PCA-DAC target diagnostics. The tolerance is $s_{\max}\max(m,n)\epsilon_{\mathrm{float32}}$ for the DAC output-projection stack. Latent errors are computed after projecting cached held-out summed-codebook embeddings to 72 PCs and reconstructing them to 1024 dimensions.}
\label{tab:pca_diagnostics}
\end{table}

\begin{figure}[t]
\centering
\includegraphics[width=\linewidth]{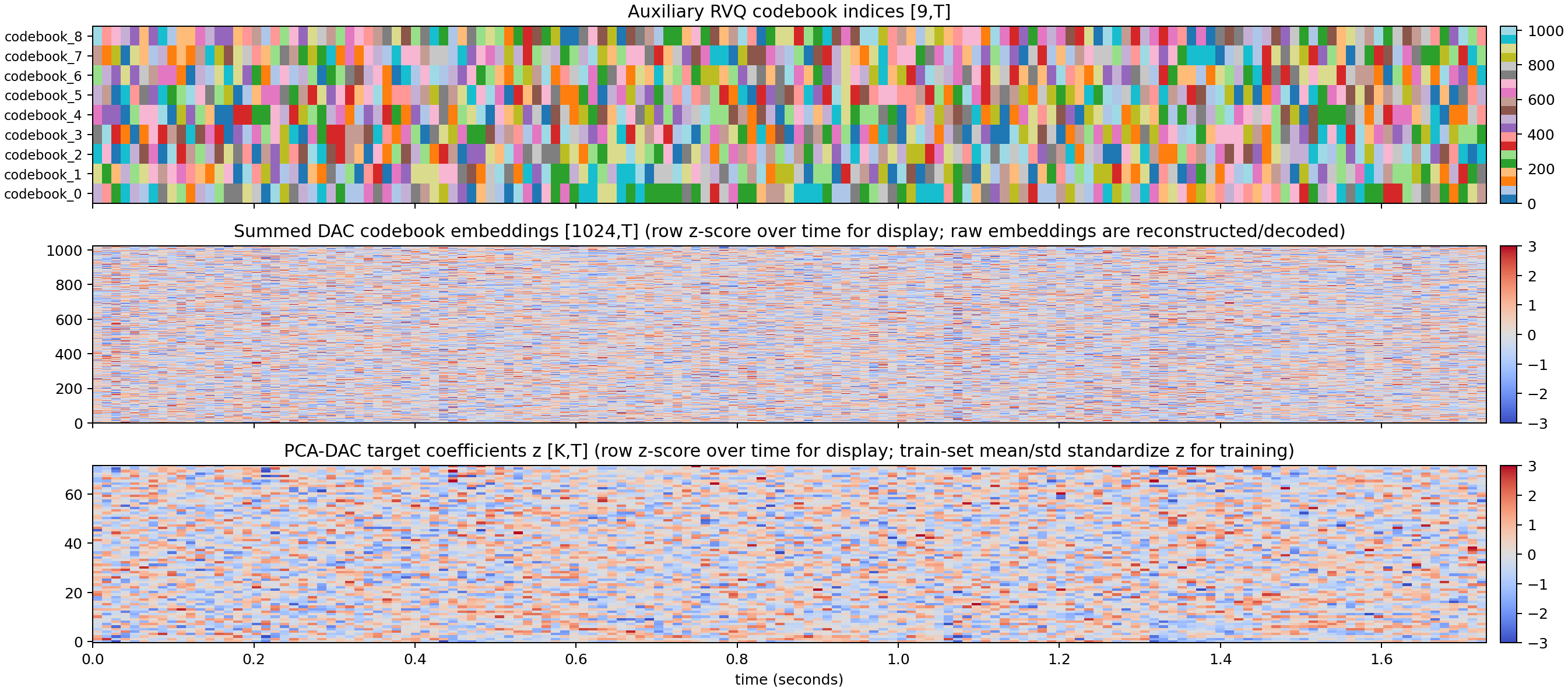}
\caption{Target-side representation inspection. RVQ codebook indices are shown as raw integer IDs for the auxiliary loss. The summed DAC embedding and PCA coefficient panels are row z-scored over time for visualization only. The continuous model target is the PCA-DAC coefficient trajectory; during training these coefficients are standardized with training-set per-dimension mean and standard deviation, then mapped back through the PCA inverse to the summed DAC embedding trajectory before waveform decoding.}
\label{fig:target_representations}
\end{figure}

\Cref{fig:target_representations} shows the target-side representations for the same held-out window. The auxiliary RVQ-codebook indices are included only to contextualize the optional RVQ-CE regularizer. The modeled object is the continuous PCA-DAC coefficient trajectory; decoding first reconstructs the full 1024-dimensional summed DAC embedding trajectory and then applies the frozen DAC decoder.

\subsection{Conditional Diffusion}

Let $x_0=\tilde{z}_{1:T}$. The forward diffusion process is
\begin{equation}
q(x_n \given x_0)
=
\mathcal{N}\!\left(
x_n;
\sqrt{\bar{\alpha}_n}x_0,
(1-\bar{\alpha}_n)I
\right),
\end{equation}
where $\bar{\alpha}_n=\prod_{i=1}^{n}\alpha_i$. A Transformer denoiser receives the noised trajectory, the diffusion index, and the seconds-aligned conditioning sequence $h_{1:T}$, and predicts the injected noise:
\begin{equation}
\hat{\epsilon}=\epsilon_\theta(x_n,n,h).
\end{equation}
The training objective is the standard noise-prediction loss,
\begin{equation}
\mathcal{L}_{\epsilon}
=
\E_{x_0,n,\epsilon}
\left[
\left\|
\epsilon-\epsilon_\theta(x_n,n,h)
\right\|_2^2
\right].
\end{equation}
where $n$ is sampled uniformly from $\{1,\dots,N\}$ and $\epsilon \sim \mathcal{N}(0,I)$. In this formulation, the model learns a conditional prior over normalized PCA trajectories rather than over waveform samples.

The implementation uses a cosine noise schedule and a DDPM-style reverse process in the normalized PCA space. Reported checkpoints vary the number of diffusion steps directly, with $N \in \{6,12,25,50\}$ for plain diffusion and $N \in \{6,12,25\}$ for the auxiliary RVQ-CE runs. During sampling, reconstructed normalized $x_0$ estimates are clipped to $[-6,6]$ before the posterior update. No classifier-free guidance is used for the quantitative results: conditioning dropout is disabled and the evaluation guidance scale is 1.0.

\subsection{Auxiliary Codec-Structure Supervision}

We additionally test whether a continuous latent denoiser benefits from supervision aligned with the discrete structure of the underlying codec. The auxiliary loss first converts the denoiser's current $\hat{x}_0$ estimate back through de-standardization and PCA inversion to a full DAC latent estimate $\hat{y}_j$. For frame $j$ and codebook $k$, define the residual against previously selected target codebook embeddings as
\[
r_{j,k}=\hat{y}_j-\sum_{k'<k} e_{k',q_{j,k'}} .
\]
The class logit for codebook entry $m$ is the negative Euclidean distance
\[
\ell_{j,k,m}=-\left\|r_{j,k}-e_{k,m}\right\|_2 ,
\]
with no learned temperature. The cross-entropy is computed over all valid frames and all RVQ codebooks using the target labels $q_{j,k}$. Gradients pass through $\hat{x}_0$, de-standardization, and PCA inversion into the denoiser; codebook embeddings and target residual terms are fixed. Because $\hat{x}_0$ is derived from the sampled training diffusion index $n$, this auxiliary term is evaluated for each training timestep sampled by the standard diffusion loss, but it is not used during inference. The resulting objective is
\begin{equation}
\mathcal{L}
=
\mathcal{L}_{\epsilon}
+\lambda_{\mathrm{ce}}\mathcal{L}_{\mathrm{ce}}.
\end{equation}
In experiments with auxiliary supervision, we use $\lambda_{\mathrm{ce}}=0.10$. This term is used only during training and does not change the sampled object: inference still generates continuous PCA-DAC trajectories, which are deterministically mapped back to full DAC latent space before decoding.

\subsection{Inference and Decoding}

At inference time, reverse diffusion samples a normalized PCA trajectory, which is then mapped back to full codec space via
\begin{align}
z_j &= s \odot \tilde{z}_j + m,\\
\hat{y}_j &= U_K z_j + \mu.
\end{align}
The waveform estimate is obtained as $\hat{a}=D_{\mathrm{DAC}}(\hat{y}_{1:T})$. Sampling therefore proceeds through a stochastic continuous latent-generation stage followed by deterministic reconstruction and codec decoding.

The stored quantitative diffusion exports use one generated sample per conditioning input. They use guidance scale 1.0 and the PyTorch sampling RNG state of the export run; metrics are averaged over clips, not over multiple stochastic samples per clip. The confidence intervals reported below therefore quantify paired clip variation only. They should not be read as including sampling-seed variability.

\begin{algorithm}[t]
\caption{Inference with \method{}}
\begin{algorithmic}[1]
\State Build seconds-aligned conditioning $h_{1:T}$ from symbolic control $c$.
\State Initialize Gaussian noise $x_N$ in normalized PCA space.
\For{$n=N,N-1,\dots,1$}
  \State Predict $\hat{\epsilon}=\epsilon_\theta(x_n,n,h)$.
  \State Apply one reverse-diffusion update.
\EndFor
\State Denormalize PCA coefficients and reconstruct full DAC latents.
\State Decode reconstructed DAC latents to waveform audio.
\end{algorithmic}
\end{algorithm}

\section{Experimental Setup}

\subsection{Data Preprocessing}

Examples are built from Groove MIDI Dataset-derived drum performances with paired MIDI and audio. For each source performance, the cache uses the original audio stream for beat analysis and target encoding. Beat times are either read from precomputed analysis metadata or estimated with Madmom \citep{boeck2016madmom}: an \texttt{RNNBeatProcessor} produces beat activations and a \texttt{DBNBeatTrackingProcessor} converts them to beat times at 100~fps, with correction disabled and a 40--320~BPM search range. These detected beat times define the four-beat segment boundaries used throughout training, plotting, and evaluation.

The cache extracts non-overlapping four-beat windows with a four-beat hop. Windows with no symbolic drum activity are discarded, and boundary-start mismatches between symbolic hits and audio-onset evidence are filtered using a 60~ms boundary window. Audio-onset evidence is computed from a short-time onset envelope with a 10~ms analysis window and 5~ms hop. The retained windows are rendered into an eight-family drum representation with state velocity, onset velocity, onset count, and family articulation IDs. These symbolic features are stored on a 250~Hz seconds grid and later sampled by the frontend at codec-frame times.

Audio is encoded with the frozen DAC model after segmentation. The diffusion cache stores the framewise summed DAC codebook-embedding trajectory, the corresponding RVQ indices for auxiliary supervision, and normalized PCA coordinates. PCA is fit on training frames only, and the validation/test frames are projected with that fixed training-set basis and normalization.

\subsection{Data and Splits}

The dataset comprises 11{,}523 training, 1{,}534 validation, and 1{,}733 test windows. Audio is encoded with the Hugging Face \texttt{descript/dac\_44khz} checkpoint of DAC \citep{kumar2023dac} at 44.1~kHz, and 9 quantizers, yielding a codec frame rate of 86.1328~Hz. The PCA target retains 72 dimensions and is fit on training frames only. Split definitions, preprocessing metadata, and hashes are versioned with the evaluation artifacts.

\subsection{Compared Systems}

The main comparison is organized around six system families:
\begin{enumerate}
  \item Reconstruction ceilings: target DAC reconstruction and target PCA reconstruction.
  \item Symbolic rendering: a deterministic renderer from the drum grid.
  \item Source-code decode and symbolic nearest-neighbor retrieval sanity baselines.
  \item Direct PCA regression: a deterministic sequence regressor with the same PCA target family.
  \item Plain PCA diffusion: denoising-step evaluations at 6, 12, 25, and 50 steps.
  \item PCA diffusion with RVQ-CE: auxiliary-loss evaluations at 6, 12, and 25 steps.
\end{enumerate}

The diffusion backbone uses target dimension 72, width 768, 6 Transformer layers, 8 attention heads, dropout 0.1, and seed 1234. The plain and auxiliary diffusion runs use the same main architecture so that the auxiliary loss is the central comparison.

The symbolic grid renderer is not MIDI playback from an external sampler in this artifact. It is a deterministic procedural renderer driven by the same eight-family grid used for conditioning: event times come from the 250~Hz seconds grid, velocities come from onset/state lanes, articulation IDs affect simple class-specific synthesis choices, and the output is written at 44.1~kHz with the same segment durations as the generated clips. The symbolic nearest-neighbor baseline retrieves from the training split only. Its query and index features are flattened 16th-grid onset counts, onset velocities, state velocities, scaled family articulation IDs, and BPM; retrieval uses normalized dot product. The implementation does not explicitly exclude the same source performance beyond the train/test split boundary. The retrieved output is the DAC decode of the selected train clip's stored source-code sequence.

\subsection{Training and Model Selection}

All learned systems are trained for 150 epochs with AdamW, learning rate $10^{-4}$, weight decay $10^{-4}$, batch size 4, and validation batch size 4. Diffusion checkpoints are selected by validation loss. The direct PCA baseline uses the same seconds-aware conditioning frontend and PCA target, but replaces the stochastic denoising objective with a deterministic Transformer regressor trained with Huber loss ($\beta=0.25$). It uses width 1024, 6 Transformer layers, 8 attention heads, dropout 0.1, and the same random seed. Thus the comparison is target-matched and conservative with respect to deterministic-regressor capacity, but it is not architecture-identical.

\subsection{Metrics}

Evaluation in generated audio is intrinsically multi-criteria, so we report \fadinf{}, \fadrtwo{}, mel MAE, broadband and band-limited onset-flux cosine, band-balance error, centroid MAE, raw RMS level MAE, crest-factor MAE, MRSTFT log-magnitude $L_1$, phase-sensitive waveform $L_1$, and real-time factor. The selected metrics are intended to expose different failure modes rather than to define a single leaderboard:
\begin{itemize}
  \item \textbf{Distributional embedding distance.} \fadinf{} compares generated and reference sets in a learned audio-embedding space, so it is useful for detecting gross distribution shift but does not verify that a specific grid produced its paired target clip. For example, symbolic nearest-neighbor retrieval has low \fadinf{} (0.025), close to the diffusion rows, while its paired mel MAE and onset-flux cosine are poor. \fadrtwo{} reports whether the repeated \fadinf{} measurements are well explained by the extrapolation fit; low values for the symbolic renderer and direct regressor warn against over-interpreting their FAD point estimates.
  \item \textbf{Paired spectral fidelity.} Mel MAE and MRSTFT log-magnitude $L_1$ compare each generated clip with its matched reference in time-frequency space. Mel MAE is the main broad spectral-distance measure; MRSTFT adds a multi-resolution view of spectral error. In the reported table, RVQ-CE at 12 steps gives the best learned-system mel MAE (5.39), while RVQ-CE at 25 steps gives the best learned-system MRSTFT point estimate (0.104).
  \item \textbf{Transient and control fidelity.} Broadband and band-limited onset-flux cosine measure whether energy changes occur at the right times, which is especially important for drums because a sample can have plausible timbre while missing or smearing controlled hits. The band-limited versions separate low-, mid-, and high-frequency transient agreement; RVQ-CE improves these values at matched step counts, and the 6-step RVQ-CE row gives the best learned-system broadband onset-flux point estimate (0.866).
  \item \textbf{Spectral balance and level.} Band-balance error, centroid MAE, raw RMS level MAE, and crest-factor MAE expose timbral and dynamic trade-offs that mel or FAD alone can hide. Raw RMS MAE is the absolute difference between clip-level RMS values in dBFS after an $-80$~dB floor over the masked four-beat region. It is a level-offset diagnostic, not a reconstruction-ceiling error: constant gain differences can produce large raw RMS values even when waveform shape, mel, and MRSTFT are nearly identical. The evaluation code now also emits a peak-normalized RMS MAE for future tables; the stored table below predates that column, so claims are not based on raw RMS.
  \item \textbf{Sample agreement and efficiency.} Waveform $L_1$ measures direct sample-level agreement and is phase-sensitive, so it is reported as a reconstruction proxy rather than as the main perceptual criterion. The direct regressor has the best learned-system waveform $L_1$ despite weaker spectral and transient metrics. RTF measures practical inference speed in the stored export run; because RVQ-CE is training-only and does not change the inference architecture, RTF differences between plain and RVQ-CE rows should be read as implementation/runtime measurements.
\end{itemize}
FAD was introduced as an unpaired distributional metric for audio evaluation \citep{kilgour2019fad}. In the reported artifacts, FAD is computed with the \texttt{clap-laion-music} embedding model from LAION-CLAP \citep{wu2023laionclap}; following recent work on FAD for generated music, we report \fadinf{} with eight repeated extrapolation runs and keep paired acoustic claims on clip-level metrics \citep{gui2024fad}. \fadrtwo{} is included to expose whether the finite-sample FAD curve is well explained by the extrapolation model; it is not treated as a perceptual quality score. MRSTFT log-magnitude $L_1$ follows the common multi-resolution STFT loss formulation used in neural vocoding work \citep{yamamoto2020parallelwavegan}.

\subsection{Statistical Testing}

Paired clip-level confidence intervals use 2{,}000 percentile bootstrap resamples over the 1{,}733 shared test indices at 95\% confidence. Paired significance tests use 2{,}000 two-sided sign-flip permutation samples over the same shared indices. For each paired metric, the best point-estimate row is compared against the remaining rows and the resulting $p$-values are Holm-adjusted within that metric. These intervals and tests cover clip variation for the single stored generated sample per conditioning input; they do not include diffusion sampling-seed variability. \fadinf{} uncertainty reflects the eight repeated extrapolation runs and is not treated as a paired significance test.

\section{Results and Analysis}

The results are organized around the supported findings rather than a raw leaderboard. All systems are evaluated on the same 1{,}733 test clips, so clip-level metrics are backed by paired bootstrap confidence intervals and paired sign-flip permutation tests where a claim depends on a paired comparison. \Cref{tab:main_results,tab:band_metrics,tab:secondary_metrics} give the aggregate point estimates. \fadinf{} is kept as a distributional metric in the aggregate table, and \fadrtwo{} is included as an extrapolation-fit diagnostic; neither is part of the paired inferential tests.

\begin{table*}[t]
\centering
\small
\setlength{\tabcolsep}{2.5pt}
\resizebox{\textwidth}{!}{%
\begin{tabular}{llrrrrrrrr}
\toprule
System & Type & Clips & \fadinf{} $\downarrow$ & \fadrtwo{} $\uparrow$ & Mel $\downarrow$ & Flux $\uparrow$ & MRSTFT $\downarrow$ & Audio $L_1$ $\downarrow$ & RTF $\downarrow$ \\
\midrule
Target DAC reconstruction & ceiling & 1733 & 0.016 & 0.848 & 0.10 & 0.999 & 0.001 & 0.0002 & 0.015 \\
Target PCA reconstruction & ceiling & 1733 & 0.016 & 0.847 & 0.10 & 0.999 & 0.001 & 0.0002 & 0.015 \\
Symbolic grid render & baseline & 1733 & 0.551 & 0.147 & 19.27 & 0.763 & 0.264 & 0.0859 & 0.004 \\
Source-code decode & sanity & 1733 & 0.016 & 0.848 & 0.10 & 0.999 & 0.001 & 0.0002 & 0.015 \\
Symbolic NN retrieval & retrieval & 1733 & 0.025 & 0.846 & 17.54 & 0.330 & 0.208 & 0.0671 & 0.018 \\
Direct PCA regressor & direct & 1733 & 0.355 & 0.169 & 13.04 & 0.836 & 0.136 & 0.0451 & 0.021 \\
PCA diffusion, 6 steps & diffusion & 1733 & 0.023 & 0.851 & 6.39 & 0.843 & 0.108 & 0.0534 & 0.060 \\
PCA diffusion, 12 steps & diffusion & 1733 & 0.021 & 0.835 & 5.75 & 0.850 & 0.106 & 0.0533 & 0.065 \\
PCA diffusion, 25 steps & diffusion & 1733 & 0.019 & 0.891 & 5.69 & 0.848 & 0.106 & 0.0536 & 0.077 \\
PCA diffusion, 50 steps & diffusion & 1733 & 0.024 & 0.744 & 5.71 & 0.839 & 0.109 & 0.0544 & 0.107 \\
PCA diffusion+RVQ-CE, 6 steps & diffusion+CE & 1733 & 0.022 & 0.830 & 5.47 & 0.866 & 0.105 & 0.0509 & 0.027 \\
PCA diffusion+RVQ-CE, 12 steps & diffusion+CE & 1733 & 0.020 & 0.845 & 5.39 & 0.864 & 0.106 & 0.0515 & 0.046 \\
PCA diffusion+RVQ-CE, 25 steps & diffusion+CE & 1733 & 0.020 & 0.848 & 5.47 & 0.863 & 0.104 & 0.0512 & 0.072 \\
\bottomrule
\end{tabular}
}
\caption{Automatically aggregated test-set metrics from stored evaluation outputs. All rows have paired acoustic metrics and \fadinf{} computed with the \texttt{clap-laion-music} embedding model. \fadrtwo{} is the coefficient of determination of the repeated \fadinf{} extrapolation fit and is reported as a diagnostic rather than as an audio-quality metric. Reconstruction ceilings and source-code decode are included as sanity references; lower is better except for onset-flux cosine and \fadrtwo{}.}
\label{tab:main_results}
\end{table*}

\begin{table}[t]
\centering
\small
\begin{tabular}{lrrrr}
\toprule
System & Low flux $\uparrow$ & Mid flux $\uparrow$ & High flux $\uparrow$ & Band bal. $\downarrow$ \\
\midrule
Symbolic grid render & 0.646 & 0.692 & 0.740 & 0.082 \\
Direct PCA regressor & 0.661 & 0.805 & 0.799 & 0.169 \\
PCA diffusion, 6 steps & 0.684 & 0.790 & 0.849 & 0.042 \\
PCA diffusion, 12 steps & 0.703 & 0.805 & 0.855 & 0.035 \\
PCA diffusion, 25 steps & 0.708 & 0.807 & 0.850 & 0.034 \\
PCA diffusion, 50 steps & 0.694 & 0.791 & 0.841 & 0.039 \\
PCA diffusion+RVQ-CE, 6 steps & 0.743 & 0.816 & 0.868 & 0.040 \\
PCA diffusion+RVQ-CE, 12 steps & 0.739 & 0.813 & 0.866 & 0.033 \\
PCA diffusion+RVQ-CE, 25 steps & 0.738 & 0.823 & 0.861 & 0.033 \\
\bottomrule
\end{tabular}
\caption{Band-limited onset-flux cosine and band-balance error for the main controllable-rendering systems. RVQ-CE improves low-, mid-, and high-band transient agreement relative to the plain diffusion variants at comparable step counts.}
\label{tab:band_metrics}
\end{table}

\begin{table}[t]
\centering
\small
\resizebox{\textwidth}{!}{%
\begin{tabular}{lrrr}
\toprule
System & Centroid MAE (Hz) $\downarrow$ & Raw RMS level MAE (dB) $\downarrow$ & Crest MAE (dB) $\downarrow$ \\
\midrule
Target DAC reconstruction & 11.1 & 7.288 & 0.008 \\
Target PCA reconstruction & 11.1 & 7.288 & 0.008 \\
Symbolic grid render & 2119.0 & 9.826 & 2.809 \\
Source-code decode & 11.1 & 7.288 & 0.008 \\
Symbolic NN retrieval & 1122.0 & 7.646 & 2.392 \\
Direct PCA regressor & 1723.1 & 5.532 & 8.309 \\
PCA diffusion, 6 steps & 546.5 & 6.749 & 1.696 \\
PCA diffusion, 12 steps & 408.8 & 6.758 & 1.626 \\
PCA diffusion, 25 steps & 394.0 & 6.964 & 1.606 \\
PCA diffusion, 50 steps & 397.9 & 6.604 & 1.729 \\
PCA diffusion+RVQ-CE, 6 steps & 408.3 & 6.770 & 1.613 \\
PCA diffusion+RVQ-CE, 12 steps & 352.8 & 6.754 & 1.595 \\
PCA diffusion+RVQ-CE, 25 steps & 364.6 & 6.422 & 1.716 \\
\bottomrule
\end{tabular}
}
\caption{Secondary paired acoustic metrics from the same completed artifact set. Raw RMS level MAE is a dBFS level-offset diagnostic and should not be interpreted as a reconstruction-ceiling error; the near-perfect reconstruction rows in the other paired metrics indicate that their large raw RMS value reflects gain handling in this stored metric rather than a failed decode.}
\label{tab:secondary_metrics}
\end{table}

\paragraph{Artifact completeness.}
The completed artifact set supports comparisons across reconstruction ceilings, symbolic baselines, direct regression, plain diffusion, and auxiliary-loss diffusion using the same metric suite. This matters for interpretation: reconstruction rows quantify codec/PCA ceiling behavior, source-code decode verifies the evaluation decode path, and symbolic nearest-neighbor retrieval separates distributional similarity from paired control fidelity. The \fadrtwo{} column is high for reconstruction/source-code rows and most diffusion rows (0.830--0.891), lower for the 50-step plain diffusion run (0.744), and weak for the symbolic grid renderer and direct regressor (0.147--0.169). A high \fadrtwo{} is only evidence that the repeated \fadinf{} extrapolation is well fit; it is not evidence that a generated clip matches its paired control target.

\paragraph{Reconstruction ceiling.}
The inferential results support treating the PCA-DAC target as a lossless-enough modeling coordinate for this experiment. Target PCA reconstruction and target DAC reconstruction have the same rounded mel MAE (0.10), onset-flux cosine (0.999), MRSTFT (0.001), and waveform $L_1$ (0.0002). The paired Target-PCA-versus-Target-DAC mel contrast is effectively zero, with a confidence interval tightly centered around zero and $p=1.0$. The source-code decode row matches the same ceiling values, so the evaluation path is consistent with the stored target representation.

\paragraph{Diffusion versus direct regression.}
The strongest supported contrast is between diffusion and deterministic PCA regression on paired spectral fidelity. Plain 25-step diffusion improves mel MAE over the direct regressor by 7.35 dB on the paired test clips (95\% CI [7.19, 7.51], $p<0.001$), and improves broadband onset-flux cosine by 0.012 (95\% CI [0.007, 0.017], $p<0.001$). Its \fadinf{} point estimate is also much closer to the reconstruction rows than the direct regressor is (0.019 versus 0.355), but this is distributional evidence rather than a paired-control test. At the same time, the direct regressor has the best learned-system phase-sensitive waveform $L_1$ in \Cref{tab:main_results}. The result is therefore a metric trade-off rather than a universal win for either family. We do not use raw RMS MAE to support this contrast because the reconstruction rows reveal a gain-handling artifact in that stored metric.

\paragraph{RVQ-CE regularization.}
The paired tests also support the usefulness of RVQ cross-entropy for efficient diffusion. At 6 denoising steps, RVQ-CE improves mel MAE by 0.92 dB (95\% CI [0.81, 1.03]) and onset-flux cosine by 0.023 (95\% CI [0.018, 0.029]) over the corresponding plain diffusion model. At 12 steps, RVQ-CE still gives a statistically resolved mel advantage of 0.36 dB (95\% CI [0.24, 0.49]) over the plain 12-step model. \Cref{tab:band_metrics} shows the same pattern in the band-limited transient metrics, where RVQ-CE improves low-, mid-, and high-band flux agreement. Because RVQ-CE is a training-only auxiliary loss, these results are best read as evidence for codec-aware regularization rather than a change in the generated representation.

\paragraph{Denoising steps.}
The step-count sweep should be read conservatively. Plain 25-step diffusion has the best plain-diffusion \fadinf{} and mel point estimate. The 50-step run is slower, has worse onset-flux agreement than the 25-step run, and has no supported mel advantage, so more reverse-diffusion steps are not justified by these results alone. All diffusion rows are faster than real time in the stored run. Because RVQ-CE is a training-only auxiliary loss and the inference architecture is otherwise the same, RTF differences should be interpreted as implementation/runtime measurements rather than architectural consequences of RVQ-CE. Runtime is therefore a secondary practical constraint rather than the main claim.

\paragraph{Baselines and secondary metrics.}
The symbolic baselines show why the paired inferential view matters. Symbolic nearest-neighbor retrieval has low \fadinf{} (0.025), close to the diffusion rows, but its paired mel MAE (17.54) and onset-flux cosine (0.330) are poor. Conversely, the symbolic renderer is very fast but substantially worse on paired spectral and transient metrics. \Cref{tab:secondary_metrics} adds another caution: raw RMS level MAE is not reliable as a reconstruction-error claim in the stored artifact, while centroid and crest-factor errors remain useful secondary views. The band-profile artifact makes the direct-regression failure mode concrete: its predicted high-band energy ratio is 0.109, compared with approximately 0.005 for target/source reconstruction and 0.005--0.007 for the diffusion rows. Among learned systems, RVQ-CE at 12 steps gives the best centroid and crest-factor point estimates, while RVQ-CE at 25 steps gives the best MRSTFT point estimate. The full artifact keeps the complete interval and significance CSVs; the manuscript reports only the claim-level contrasts needed to support the discussion.

\section{Conclusion}

We presented \method{}, a seconds-aligned PCA-DAC latent-diffusion model for explicit symbolic-to-audio drum rendering. The formulation combines physical-time conditioning, a compact continuous coordinate system for DAC summed-codebook latents, deterministic reconstruction to full DAC latent space, and optional RVQ cross-entropy regularization. The complete artifact set supports three main conclusions. First, the PCA-DAC target is a practical modeling space in this configuration: target DAC and target PCA reconstruction are effectively identical at the reported precision, so the main errors come from generation rather than from the 72-dimensional PCA bottleneck. Second, PCA diffusion improves paired spectral and transient metrics over deterministic PCA regression and symbolic grid rendering, while its \fadinf{} point estimates show less distributional shift than those baselines; direct regression remains better on phase-sensitive waveform $L_1$. Third, RVQ-CE improves the most efficient diffusion settings and provides evidence that codec-aware auxiliary supervision can help continuous latent denoising.

The broader implication is methodological: for controlled drum rendering, a continuous codec-latent diffusion target can preserve an explicit symbolic interface while avoiding direct waveform modeling and avoiding discrete-token generation as the primary output. The results support \method{} as a coherent and reproducible formulation for neural drum rendering under explicit symbolic control.

\section{Code Availability}

The full training, inference, evaluation, and UI code will be released in a cleaned public repository after packaging is finalized. The release will include the scripts used to regenerate the reported tables, representation figures, and qualitative UI examples.

\section{Limitations}

This study is intentionally narrow. It evaluates short four-beat drum-rendering windows rather than full musical arrangements, and the conditioning input is an explicit drum grid rather than open text, audio prompts, or reference-timbre examples. The PCA representation is linear, fixed, and tied to a specific DAC configuration; this improves reproducibility and interpretability but does not establish that the same dimensionality or representation is optimal for other codecs, longer windows, or non-drum audio. The evaluation is automatic and artifact-backed, but it is not a human listening study, so claims about perceived musical quality must remain cautious \citep{groetschla2025benchmarking}. Waveform $L_1$ and spectral/transient metrics also disagree in important cases, which limits any single-metric interpretation of model quality. The diffusion evaluation uses one generated sample per test clip, so the reported confidence intervals do not measure sampling-seed variability. The stored raw RMS level metric also contains a gain-handling artifact for reconstruction rows, so RMS-based claims are withheld until the peak-normalized RMS column is regenerated. Finally, the interactive UI is useful for qualitative inspection and demos, but it is not treated as user-study evidence.


\begin{thebibliography}{99}

\bibitem[Abid et~al.(2019)]{abid2019gradio}
Abubakar Abid, Abdalla Abdalla, Ali Abid, Dawood Khan, Abdulrahman Alfozan, and James Zou.
\newblock Gradio: Hassle-free sharing and testing of ML models in the wild.
\newblock \emph{arXiv preprint arXiv:1906.02569}, 2019.

\bibitem[Agostinelli et~al.(2023)]{agostinelli2023musiclm}
Andrea Agostinelli, Timo I. Denk, Zal\'an Borsos, Jesse Engel, Mauro Verzetti, Antoine Caillon, Qingqing Huang, Aren Jansen, Adam Roberts, Marco Tagliasacchi, Matthew Sharifi, Neil Zeghidour, and Christian Frank.
\newblock MusicLM: Generating music from text.
\newblock \emph{arXiv preprint arXiv:2301.11325}, 2023.

\bibitem[B\"ock et~al.(2016)]{boeck2016madmom}
Sebastian B\"ock, Filip Korzeniowski, Jan Schl\"uter, Florian Krebs, and Gerhard Widmer.
\newblock madmom: A new Python audio and music signal processing library.
\newblock In \emph{Proceedings of the 24th ACM International Conference on Multimedia}, pages 1174--1178, 2016.

\bibitem[Borsos et~al.(2023)]{borsos2023audiolm}
Zal\'an Borsos, Rapha\"el Marinier, Damien Vincent, Eugene Kharitonov, Olivier Pietquin, Matt Sharifi, Olivier Teboul, David Grangier, Marco Tagliasacchi, and Neil Zeghidour.
\newblock AudioLM: A language modeling approach to audio generation.
\newblock \emph{IEEE/ACM Transactions on Audio, Speech, and Language Processing}, 31:2523--2533, 2023.

\bibitem[Brosnan(2026)]{brosnan2026darc}
Trey Brosnan.
\newblock DARC: Drum accompaniment generation with fine-grained rhythm control.
\newblock \emph{arXiv preprint arXiv:2601.02357}, 2026.

\bibitem[Chen et~al.(2024)]{chen2024pixartalpha}
Junsong Chen, Jincheng Yu, Chongjian Ge, Lewei Yao, Enze Xie, Yue Wu, Zhongdao Wang, James Kwok, Ping Luo, Huchuan Lu, and Zhenguo Li.
\newblock PixArt-$\alpha$: Fast training of diffusion transformer for photorealistic text-to-image synthesis.
\newblock In \emph{Proceedings of ICLR}, 2024.

\bibitem[Cui et~al.(2026)]{cui2026breakthebeat}
Shuyang Cui, Zhi Zhong, Qiyu Wu, Zachary Novack, Woosung Choi, Keisuke Toyama, Kin Wai Cheuk, Junghyun Koo, Yukara Ikemiya, Christian Simon, Chihiro Nagashima, and Shusuke Takahashi.
\newblock Break-the-Beat! Controllable MIDI-to-drum audio synthesis.
\newblock In \emph{Proceedings of ICASSP}, 2026.
\newblock doi:10.1109/ICASSP55912.2026.11464812.

\bibitem[Defossez et~al.(2022)]{defossez2022encodec}
Alexandre Defossez, Jade Copet, Gabriel Synnaeve, and Yossi Adi.
\newblock High fidelity neural audio compression.
\newblock \emph{arXiv preprint arXiv:2210.13438}, 2022.

\bibitem[Dhariwal and Nichol(2021)]{dhariwal2021diffusion}
Prafulla Dhariwal and Alexander Nichol.
\newblock Diffusion models beat GANs on image synthesis.
\newblock In \emph{Advances in Neural Information Processing Systems}, 2021.

\bibitem[Evans et~al.(2024a)]{evans2024stableaudio}
Zach Evans, CJ Carr, Josiah Taylor, Scott H. Hawley, and Jordi Pons.
\newblock Fast timing-conditioned latent audio diffusion.
\newblock In \emph{Proceedings of the 41st International Conference on Machine Learning}, 2024.

\bibitem[Evans et~al.(2024b)]{evans2024longform}
Zach Evans, Julian D. Parker, CJ Carr, Zack Zukowski, Josiah Taylor, and Jordi Pons.
\newblock Long-form music generation with latent diffusion.
\newblock \emph{arXiv preprint arXiv:2404.10301}, 2024.

\bibitem[Gillick et~al.(2019)]{gillick2019learning}
Jon Gillick, Adam Roberts, Jesse Engel, Douglas Eck, and David Bamman.
\newblock Learning to groove with inverse sequence transformations.
\newblock In \emph{Proceedings of the International Conference on Machine Learning}, 2019.

\bibitem[Google Magenta(2019)]{magenta2019groove}
Google Magenta.
\newblock The Groove MIDI Dataset.
\newblock \url{https://magenta.tensorflow.org/datasets/groove}, 2019.

\bibitem[Gr\"otschla et~al.(2025)]{groetschla2025benchmarking}
Florian Gr\"otschla, Ahmet Solak, Luca A. Lanzend\"orfer, and Roger Wattenhofer.
\newblock Benchmarking music generation models and metrics via human preference studies.
\newblock In \emph{Proceedings of ICASSP}, 2025.

\bibitem[Gui et~al.(2024)]{gui2024fad}
Azalea Gui, Hannes Gamper, Sebastian Braun, and Dimitra Emmanouilidou.
\newblock Adapting Fr{\'e}chet Audio Distance for generative music evaluation.
\newblock In \emph{Proceedings of ICASSP}, 2024.

\bibitem[Hawthorne et~al.(2022)]{hawthorne2022multiinstrument}
Curtis Hawthorne, Ian Simon, Ryan Swavely, Ethan Manilow, and Jesse Engel.
\newblock Multi-instrument music synthesis with spectrogram diffusion.
\newblock In \emph{Proceedings of ISMIR}, 2022.

\bibitem[Ho et~al.(2020)]{ho2020ddpm}
Jonathan Ho, Ajay Jain, and Pieter Abbeel.
\newblock Denoising diffusion probabilistic models.
\newblock In \emph{Advances in Neural Information Processing Systems}, 2020.

\bibitem[Jajoria and McDermott(2024)]{jajoria2024text}
Pushkar Jajoria and James McDermott.
\newblock Text-conditioned symbolic drumbeat generation using latent diffusion models.
\newblock \emph{arXiv preprint arXiv:2408.02711}, 2024.

\bibitem[Kilgour et~al.(2019)]{kilgour2019fad}
Kevin Kilgour, Mauricio Zuluaga, Dominik Roblek, and Matthew Sharifi.
\newblock Fr{\'e}chet Audio Distance: A reference-free metric for evaluating music enhancement algorithms.
\newblock In \emph{Proceedings of Interspeech}, 2019.

\bibitem[Kumar et~al.(2023)]{kumar2023dac}
Rithesh Kumar, Prem Seetharaman, Alejandro Luebs, Ishaan Kumar, and Kundan Kumar.
\newblock High-fidelity audio compression with improved RVQGAN.
\newblock In \emph{Advances in Neural Information Processing Systems}, 2023.

\bibitem[Liu et~al.(2024)]{liu2024audioldm2}
Haohe Liu, Yi Yuan, Xubo Liu, Xinhao Mei, Qiuqiang Kong, Qiao Tian, Yuping Wang, Wenwu Wang, Yuxuan Wang, and Mark D. Plumbley.
\newblock AudioLDM 2: Learning holistic audio generation with self-supervised pretraining.
\newblock \emph{IEEE/ACM Transactions on Audio, Speech, and Language Processing}, 32:2871--2883, 2024.

\bibitem[Nistal et~al.(2020)]{nistal2020drumgan}
Javier Nistal, Stefan Lattner, and Ga\"el Richard.
\newblock DrumGAN: Synthesis of drum sounds with timbral feature conditioning using generative adversarial networks.
\newblock In \emph{Proceedings of ISMIR}, 2020.

\bibitem[O'Reilly et~al.(2025)]{oreilly2025tria}
Patrick O'Reilly, Julia Barnett, Hugo Flores Garc\'ia, Annie Chu, Nathan Pruyne, Prem Seetharaman, and Bryan Pardo.
\newblock The Rhythm In Anything: Audio-prompted drums generation with masked language modeling.
\newblock In \emph{Proceedings of ISMIR}, 2025.

\bibitem[Peebles and Xie(2023)]{peebles2023dit}
William Peebles and Saining Xie.
\newblock Scalable diffusion models with transformers.
\newblock In \emph{Proceedings of the IEEE/CVF International Conference on Computer Vision}, pages 4195--4205, 2023.

\bibitem[Preechakul et~al.(2022)]{preechakul2022diffae}
Konpat Preechakul, Nattanat Chatthee, Suttisak Wizadwongsa, and Supasorn Suwajanakorn.
\newblock Diffusion autoencoders: Toward a meaningful and decodable representation.
\newblock In \emph{Proceedings of the IEEE/CVF Conference on Computer Vision and Pattern Recognition}, 2022.

\bibitem[Rombach et~al.(2022)]{rombach2022ldm}
Robin Rombach, Andreas Blattmann, Dominik Lorenz, Patrick Esser, and Bj\"orn Ommer.
\newblock High-resolution image synthesis with latent diffusion models.
\newblock In \emph{Proceedings of the IEEE/CVF Conference on Computer Vision and Pattern Recognition}, 2022.

\bibitem[Rouard and Hadjeres(2021)]{rouard2021crash}
Simon Rouard and Ga\"etan Hadjeres.
\newblock CRASH: Raw audio score-based generative modeling for controllable high-resolution drum sound synthesis.
\newblock In \emph{Proceedings of ISMIR}, 2021.

\bibitem[Soiledis et~al.(2026)]{soiledis2026drumsynthesis}
Konstantinos Soiledis, Maximos Kaliakatsos-Papakostas, Dimos Makris, and Konstantinos Tsamis.
\newblock Drum synthesis from expressive drum grids via neural audio codecs.
\newblock \emph{arXiv preprint arXiv:2605.10281}, 2026.

\bibitem[Song et~al.(2024)]{song2024pcaddpm}
Myung Keun Song, Asim Niaz, Muhammad Umraiz, Ehtesham Iqbal, Shafiullah Soomro, and Kwang Nam Choi.
\newblock Denoising diffusion-based image generation model using principal component analysis.
\newblock \emph{IEEE Access}, 12:170487--170498, 2024.

\bibitem[Tang et~al.(2025)]{tang2025midivalle}
Jiatong Tang, Xin Wang, Zhenyu Zhang, Junichi Yamagishi, Geraint Wiggins, and Gy\"orgy Fazekas.
\newblock MIDI-VALLE: Improving expressive piano performance synthesis through neural codec language modelling.
\newblock In \emph{Proceedings of ISMIR}, 2025.

\bibitem[Wu et~al.(2023)]{wu2023laionclap}
Yusong Wu, Ke Chen, Tianyu Zhang, Yuchen Hui, Taylor Berg-Kirkpatrick, and Shlomo Dubnov.
\newblock Large-scale contrastive language-audio pretraining with feature fusion and keyword-to-caption augmentation.
\newblock In \emph{Proceedings of ICASSP}, 2023.

\bibitem[Yamamoto et~al.(2020)]{yamamoto2020parallelwavegan}
Ryuichi Yamamoto, Eunwoo Song, and Jae-Min Kim.
\newblock Parallel WaveGAN: A fast waveform generation model based on generative adversarial networks with multi-resolution spectrogram.
\newblock In \emph{Proceedings of ICASSP}, 2020.

\bibitem[Zeghidour et~al.(2022)]{zeghidour2022soundstream}
Neil Zeghidour, Alejandro Luebs, Ahmed Omran, Jan Skoglund, and Marco Tagliasacchi.
\newblock SoundStream: An end-to-end neural audio codec.
\newblock \emph{IEEE/ACM Transactions on Audio, Speech, and Language Processing}, 30:495--507, 2022.

\end{thebibliography}
\end{document}